\documentclass[pre,twocolumn,amsmath,amssymb,nofootinbib,floatfix,superscriptaddress]{revtex4}

\usepackage{graphicx,bm,xcolor, bbold}
\usepackage{enumerate}
\usepackage{graphicx,bm}
\usepackage[normalem]{ulem} 

\makeatletter
\def\graphicscale{\twocolumn@sw{0.4}{0.4}}
\def\graphicthreescale{\twocolumn@sw{0.3}{0.4}}

\begin{document}

\title{Out-of-equilibrium critical dynamics of the three-dimensional \\
  ${\mathbb Z}_2$ gauge model along critical relaxational flows}

\author{Claudio Bonati} 
\affiliation{Dipartimento di Fisica dell'Universit\`a di Pisa and INFN,
        Largo Pontecorvo 3, I-56127 Pisa, Italy}

\author{Haralambos Panagopoulos} 
\affiliation{Department of Physics, University of Cyprus,
P.O. Box 20537, 1678 Nicosia, Cyprus}

\author{Ettore Vicari} 
\affiliation{Dipartimento di Fisica dell'Universit\`a di Pisa,
        Largo Pontecorvo 3, I-56127 Pisa, Italy}

\date{\today}

\begin{abstract}
We address the out-of-equilibrium critical dynamics of the
three-dimensional lattice ${\mathbb Z}_2$ gauge model, and in
particular the critical relaxational flows arising from instantaneous
quenches to the critical point, driven by purely relaxational
(single-spin-flip Metropolis) upgradings of the link ${\mathbb Z}_2$
gauge variables. We monitor the critical relaxational dynamics by
computing the energy density, which is the simplest local
gauge-invariant quantity that can be measured in a lattice gauge
theory.  The critical relaxational flow of the three-dimensional
lattice ${\mathbb Z}_2$ gauge model is analyzed within an
out-of-equilibrium finite-size scaling framework, which allows us to
compute the dynamic critical exponent $z$ associated with the purely
relaxational dynamics of the three-dimensional ${\mathbb Z}_2$ gauge
universality class. We obtain $z=2.610(15)$, which significantly
improves earlier results obtained by other methods, in particular
those obtained by analyzing the equilibrium critical dynamics.
\end{abstract}

\maketitle


\section{Introduction}
\label{intro}

Gauge symmetries are crucial features of effective theories that
describe collective phenomena developing topological transitions,
driven by extended charged excitations with nonlocal order parameters,
or by a nontrivial interplay between long-range scalar fluctuations
and nonlocal topological gauge modes.  Examples of such transitions
are observed in lattice gauge theories with discrete ${\mathbb Z}_N$
and continuous U(1) gauge symmetries, see, e.g.,
Refs.~\cite{Sachdev-book2,BPV-24-rev,Senthil-23,Sachdev-19} for recent
reviews.  While the static critical properties in the presence of
gauge symmetries have been much investigated, see, e.g.,
Refs.~~\cite{Wegner-71,BDI-74,OS-78,FS-79,Kogut-79,Savit-80,
  DH-81,FM-83,KK-85, KK-86,
  BN-87,MS-90,LRT-93,KKS-94,HT-96,IKK-96,KKLP-98,
  SF-00,SSS-02,SSSNH-02,KNS-02,MHS-02,SM-02,NRR-03,SSNHS-03,KS-08,
  CAP-08,KNNSWS-15,NCSOS-15,
  BPV-20-hcAH,BPV-21-ncAH,SSN-21,BPV-22,BPV-22-z2h,BP-23,BF-23,ZZV-23,
  BPV-23-chgf,BPV-24-z2gaugeN,
  BPV-24-onstar,XPK-24,BPV-24-decQ2,SSN-24,BPV-24-ncAH,BPSV-24},
achieving a satisfactory understanding of the various critical
scenarios that can be realized~\cite{BPV-24-rev}, much less is known
about their critical dynamics in both equilibrium and
out-of-equilibrium conditions. We believe that extending our knowledge
of the critical scenarios to dynamic, equilibrium and
out-of-equilibrium, situations may lead to a deeper understanding of
the mechanisms underlying topological transitions in the presence of
gauge symmetries.  We recall that a fair understanding of the critical
dynamics has been already achieved in the case of classical and
quantum many-body systems without gauge symmetries, see, e.g.,
Refs.~\cite{Ma-book,HH-77,CG-05,FM-06,CEGS-12,RV-21}.

The three-dimensional (3D) ${\mathbb Z}_2$ gauge
model~\cite{Wegner-71} is a paradigmatic lattice system undergoing a
finite-temperature topological transition without any local order
parameter~\cite{Wegner-71, Sachdev-19,BPV-24-rev}, separating the
high-temperature deconfined phase from the low-temperature confined
phase.  By duality~\cite{Wegner-71,Savit-80}, this model can be
related to the standard 3D Ising model (in the absence of external
magnetic fields breaking the ${\mathbb Z}_2$ symmetry), implying that
their static critical behaviors share analogous asymptotic behaviors
for the diverging length scale $\xi$ of their critical correlations,
$\xi \sim |T-T_c|^{-\nu}$ with $\nu \approx 0.630$~\cite{PV-02}.

The equilibrium and out-of-equilibrium dynamics at continuous
transitions is generally affected by critical slowing down.  The time
scale of the critical modes diverges as $\tau\sim \xi^z$, where $z$ is
an independent dynamic exponent, whose value also depends on the type
of dynamics~\cite{HH-77,FM-06}. The simplest example is provided by
the purely relaxational dynamics associated with a stochastic Langevin
equation where only dissipative couplings are present, with no
conservation laws, usually called dynamic model
A~\cite{Ma-book,HH-77}. In lattice models the purely relaxational
dynamics can be realized by standard Metropolis upgradings of the
lattice variables~\cite{Binder-76}.

Despite the fact that 3D ${\mathbb Z}_2$ gauge and Ising spin models
share the same static critical behavior, they exhibit different
critical dynamics under local relaxational upgradings. Indeed the time
scale of their critical modes develop different power laws, as shown
by some numerical analyses of the purely relaxational dynamics, under
equilibrium and out-of-equilibrium
conditions~\cite{BKKLS-90,XCMCS-18,BPV-25}.  The most recent results
for the dynamic exponent $z$ of the purely relaxational dynamics of
the 3D ${\mathbb Z}_2$ gauge model are $z=2.55(6)$, obtained by
analyzing the critical relaxational dynamics of model in equilibrium
conditions~\cite{BPV-25}, and $z=2.70(3)$, obtained by analyzing the
out-of-equilibrium dynamics when slowly crossing the transition
point~\cite{XCMCS-18}. Although they appear slightly inconsistent with
each other, they are both definitely larger than the relaxational
dynamic exponent $z=2.0245(15)$ of the 3D Ising
model~\cite{Hasenbusch-20}.  This substantial difference can be
explained by recalling that the duality relation between the 3D
${\mathbb Z}_2$ gauge and standard Ising models is nonlocal.
Therefore local relaxational upgradings may act differently, giving
rise to different power laws associated with their respective critical
slowing down.

To further investigate the out-of-equilibrium critical dynamics at the
topological transition of the 3D lattice ${\mathbb Z}_2$ gauge model,
we report numerical analyses of the critical relaxational
flows~\cite{PV-24} arising from instantaneous quenches to the critical
point, starting from thermalized conditions within the critical
region.  Since the topological transition of the 3D ${\mathbb Z}_2$
gauge model does not have local order parameters, the choice of the
optimal observable to monitor the system during out-of-equilibrium
evolutions is not straightforward. In this paper we focus on the
simplest choice given by the local gauge-invariant energy density $E$,
more precisely the subtracted energy density $E_s=E-E_c$ where $E_c$
is the equilibrium energy density at the critical point. A crucial
point of our approach is that, contrary to what happens at
equilibrium, the time-dependent subtracted energy density $E_s$ along
the relaxational flows is not affected by the complications arising
from the presence of analytical backgrounds at the transition
point~\cite{PV-24}. This is essentially due to the fact that the
short-ranged modes responsible for the analytical background and the
long-range critical modes have substantially different relaxational
times around the critical point: the short-ranged modes are expected to
thermalize much faster than the critical modes. This fact makes it
possible to effectively isolate the out-of-equilibrium finite-size
scaling (FSS) terms of the energy density, and therefore the power law
of the critical time scale. As we shall see, this approach allows us
to compute the dynamic exponent $z$, obtaining $z=2.610(15)$, which
significantly improves earlier
results~\cite{BKKLS-90,XCMCS-18,BPV-25}.

The paper is organized as follows. In Sec.~\ref{z2prot} we introduce
the 3D ${\mathbb Z}_2$ gauge model, summarizing its static critical
properties, and present the dynamic protocol adopted to simulate the
critical relaxational flows. In Sec.~\ref{enesca} we discuss the
out-of-equilibrium behavior of the energy density along the critical
relaxational flow within an out-of-equilibrium FSS framework.  In
Sec.~\ref{numres} we report our numerical analyses based on Monte
Carlo (MC) simulations, leading to the estimate of the dynamic
exponent $z$. Finally, in Sec.~\ref{conclu} we draw our conclusions.

\section{Critical relaxational flow of  the 3D ${\mathbb Z}_2$ gauge model}
\label{z2prot}

\subsection{The model}
\label{model}

The 3D lattice ${\mathbb Z}_2$ gauge model~\cite{Wegner-71} is a
paradigmatic model undergoing a finite-temperature topological
transition~\cite{Sachdev-19,BPV-24-rev}.  Its Hamiltonian, defined on
a cubic lattice, reads
\begin{equation}
H_G = - K \sum_{{\bm x},\mu>\nu}
   \sigma_{{\bm
      x},\mu} \,\sigma_{{\bm x}+\hat{\mu},\nu} \,\sigma_{{\bm
       x}+\hat{\nu},\mu} \,\sigma_{{\bm x},\nu},
   \label{Hgz2}
\end{equation}
where $\sigma_{{\bm x},\mu}=\pm 1$ is the link variable associated with
the bond starting from site ${\bm x}$ in the positive $\mu$ direction,
$\mu=1,2,3$.  The Hamiltonian parameter $K$ plays the role of inverse
gauge coupling, therefore the $K\to\infty$ limit corresponds to the
small gauge-coupling limit. In the following we set the temperature
$T=1$, so that the partition function describing the equilibrium
static properties reads
\begin{equation}
  Z_G =\sum_{\{\sigma\}} e^{-H_G}.
  \label{partfunc}
\end{equation}
In our numerical FSS analyses we will consider cubic-like systems of
size $L$ with periodic boundary conditions.

By a nonlocal duality relation~\cite{Wegner-71,Savit-80}, the
partition function $Z_G$ in the infinite-volume limit can be exactly
related to that of the standard cubic-lattice Ising model without
external fields coupled to the spin variables, i.e.
\begin{equation}
  Z_{I} = \sum_{\{s\}} e^{-H_{I}},\qquad
  H_{I} = - J \sum_{{\bm x},\mu} s_{\bm x} \,s_{{\bm x}+\hat\mu},
  \label{His}
  \end{equation}
where $s_{{\bm x},\mu}=\pm 1$ are spin variables associated with the
sites ${\bm x}$ of the cubic lattice, and the relation between $J$ and
$K$ is 
\begin{equation}
J = -{1\over 2} {\rm ln}\,{\rm tanh}\,K.
  \label{z2gaugecrrel}
\end{equation}
Notice that this duality mapping of the partition functions does not
generally extend to finite-size systems with boundary conditions. In
particular, it does not exactly relate the partition functions of
${\mathbb Z}_2$ gauge and Ising models when both of them have periodic
boundary conditions~\cite{BPV-24-decQ2}, for which the
equivalence is only recovered in the thermodynamic limit.

\subsection{The finite-temperature topological transition}
\label{toptrans}

The 3D lattice ${\mathbb Z}_2$ gauge model undergoes a continuous
transition separating a high-$K$ deconfined phase from a low-$K$
confined phase.  Its transition point and static critical behavior can
be obtained by exploiting the duality with the Ising model
(\ref{His}), for which accurate estimates of the critical point
$J_{c}$ are known, in particular, $J_c=0.221654626(5)$~\cite{FXL-18}.
This allows us to obtain a corresponding accurate estimate of the
critical point $K_c$ of the 3D ${\mathbb Z}_2$ gauge model:
\begin{equation}
K_c = -{1\over 2} {\rm ln}\,{\rm tanh}\,J_c = 0.761413292(11).
  \label{z2gaugecr}
\end{equation}
The equivalence (up to analytical terms) of the partition functions of
the 3D ${\mathbb Z}_2$ gauge and standard Ising models also implies
that they share the same critical exponent $\nu$ related to the
divergence of the correlation length, and also the leading
scaling-correction exponent $\omega$.  In the following we use the
Ising estimates~\cite{KPSV-16} $\nu=0.629971(4)$ for the length-scale
critical exponent, $\alpha = 2 - 3\nu = 0.110087(12)$ for the
specific-heat exponent, and $\omega= 0.8297(2)$ for the exponent
associated with the leading scaling corrections (see also
Refs.~\cite{Hasenbusch-21,FXL-18,KP-17,Hasenbusch-10,CPRV-02,GZ-98}).
As we shall see below, the analogy of the asymptotic critical
behaviors of 3D ${\mathbb Z}_2$ gauge and Ising models does not extend
to the critical dynamics, essentially because the highly nonlocal
duality mapping~\cite{Wegner-71,Savit-80} does not extend to the local
relaxational dynamics of both models.

One may also obtain accurate estimates of the energy density of the
$\mathbb{Z}_2$ gauge model,
\begin{equation}
  E = - {1\over L^d} \langle   \sum_{{\bm
      x},\mu>\nu}
   \sigma_{{\bm
      x},\mu} \,\sigma_{{\bm x}+\hat{\mu},\nu} \,\sigma_{{\bm
       x}+\hat{\nu},\mu} \,\sigma_{{\bm x},\nu}\,\rangle,
   \label{defened0}
\end{equation}
by exploiting the duality mapping of the partition function $Z_G$ to
that of the 3D Ising model (\ref{His}), for which accurate numerical
results can be obtained by more effective MC algorithms, such as
cluster algorithms~\cite{Wolff-89}. In particular, one may compute the
energy density
\begin{equation}
E_I= - {1\over L^{d}} \langle \, \sum_{{\bm x},\mu} s_{\bm x}
\,s_{{\bm x}+\hat\mu}\, \rangle
\label{isiene}
\end{equation}
of the 3D Ising model (\ref{His}) at
the critical point $J_c$. Then, by applying the duality mapping,
one can determine the energy density $E_c$ at the critical point $K_c$
of the 3D ${\mathbb Z}_2$ gauge model, which we will use later in our
analyses.  For this purpose, we performed MC simulations of the Ising
system (\ref{His}) with periodic boundary conditions at the critical
point $J_c$, for various lattice sizes up to $L=300$, using mixtures
of cluster and Metropolis algorithms. To obtain the critical
infinite-volume energy density $E_{Ic}\equiv E_{I}(J_c,L\to\infty)$,
these finite-size data are extrapolated by fitting them to the
expected asymptotic large-$L$ behavior $E_{Ic} + c\,L^{-(d-1/\nu)}$,
see, e.g., Ref.~\cite{PV-02} and also Eq.~(\ref{leadINGFSSene})
below. This procedure allows us to obtain the accurate estimate
$E_{Ic} = -0.990612(7)$ (this estimate is in agreement with, and
significantly improves, earlier results, see, e.g.,
Refs.~\cite{HP-98,HPV-99}).  Then, using the duality relation of the
free-energy densities of the ${\mathbb Z}_2$ gauge and Ising
models~\cite{Wegner-71,Savit-80},~\footnote{The relation between the
free-energy densities $F_G$ and $F_I$ in the thermodynamic limit is
given by~\cite{Wegner-71,Savit-80} (up to an irrelevant additive
constant)
\begin{eqnarray}
  &&F_G(K) = - {1\over L^{d}} \ln Z_G(K) = 
  F_0(K) + F_I[J(K)], \label{FgK}\\
  &&F_0(K)=-{3\over 2} \ln\sinh(2K),\quad
  F_I(J) = - {1\over L^{d}} \ln Z_I(J),
\nonumber
\end{eqnarray}
where $J(K)$ is given in Eq.~(\ref{z2gaugecrrel}).  Then, the energy
density of the 3D ${\mathbb Z}_2$ gauge model can be obtained by
differentiating with respect to $K$,
\begin{eqnarray}
  E = {\partial F_G\over \partial K} = {\partial F_0\over \partial K}
  + E_{I}[J(K)]{\partial J(K)\over \partial K},
  \label{ene}
\end{eqnarray}
where $E_I = \partial F_I/\partial J$.  Finally, by evaluating it at
$K=K_c$, using $J_c=J(K_c)$ and the Ising critical value $E_{Ic}\equiv
E_I(J_c)$, one obtains Eq.~(\ref{ecrit}).}  we obtain the critical
infinite-volume energy density of the 3D lattice ${\mathbb Z}_2$ gauge
model with a relative accuracy of about $10^{-6}$, i.e.,
\begin{equation}
  E_c = - 2.845971(3).
  \label{ecrit}
\end{equation}

We finally mention that the area law of the Wilson loops $W_C$,
defined as the product of the link variables along a closed contour
$C$ within a plane, provides a nonlocal order parameter for the
topological transition of the 3D ${\mathbb Z}_2$ gauge
model~\cite{Wegner-71}. Indeed, their asymptotic large-size dependence
changes at the transition point $K_c$: from the small-$K$ area law
$W_C\sim \exp(- c_a A_C)$ (where $A_C$ is the area enclosed by the
contour $C$ and $c_a>0$ is a constant) to the large-$K$ perimeter law
$W_C\sim \exp(- c_p P_C)$ (where $P_C$ is the perimeter of the contour
$C$ and $c_p>0$ is a constant).

\subsection{Critical relaxational flow}
\label{relprot}

To investigate the out-of-equilibrium critical dynamics at the
topological transition, we consider {\em soft} quench protocols around
the transition point~\cite{PRV-18,RV-21}, so that the
out-of-equilibrium evolution occurs within the critical region.  In
particular, we study the out-of-equilibrium critical relaxational flow
arising from an instantaneous quench of the gauge parameter $K$, from
$K<K_c$ to $K_c$. In practice, we consider the following
protocol~\cite{PV-24}:

(i) We start from a Gibbs ensemble of equilibrium configurations at
$K<K_c$.

(ii) These configurations are the starting point for an
out-equilibrium critical relaxational flow at the critical point
$K_c$, parameterized by the dimensionless time $t$ and starting at
$t=0$.  This is achieved by making the system evolve using a purely
local relaxational Metropolis dynamics at inverse coupling
$K_c$~\cite{HH-77,Binder-76}.

Since the dynamics is purely relaxational, the critical equilibrium of
finite-size systems is eventually recovered for large times, which
tend to be larger and larger with increasing $L$, due to the critical
slowing down.

In our study we monitor the out-of-equilibrium dynamics arising from
this quench protocol by the time dependence of the gauge-invariant
energy density
\begin{equation}
  E(t) = - {1\over L^d} \big\langle
  \sum_{{\bm
      x},\mu>\nu}
   \sigma_{{\bm
      x},\mu} \,\sigma_{{\bm x}+\hat{\mu},\nu} \,\sigma_{{\bm
       x}+\hat{\nu},\mu} \,\sigma_{{\bm x},\nu}\,\big\rangle_t ,
   \label{defened}
   \end{equation}
defined as the average over the configurations obtained at a fixed
time $t$ along the relaxational flow. We do not consider nonlocal
observables, like Wilson loops or Polyakov lines, whose analysis
presents nontrivial drawbacks related to their complicated
renormalizations arising from short-ranged fluctuations, see, e.g.,
Ref.~\cite{Dorn-86}, and also a significantly larger computational
effort to obtain sufficiently accurate results.

It is worth mentioning that the critical relaxational flow at
finite-temperature transitions is similar to the so-called gradient
flow that is often exploited to numerically study the four-dimensional
lattice quantum chromodynamics (QCD), which is the theory of strong
interactions~\cite{Weinberg-book1,Weinberg-book2,Creutz-book,MM-book},
in order to define a running coupling from the lattice energy density,
see, e.g., Refs.~\cite{Luscher-10,LW-11,HN-16}.  Indeed, since the
continuum limit of lattice QCD is realized in the zero-coupling
(zero-temperature) limit, where the lattice length scale diverges
exponentially, the critical (fixed-point) relaxational flow becomes a
simple deterministic gradient flow at vanishing bare gauge coupling
(corresponding to zero temperature), which is equivalent to a Langevin
equation without stochastic term~\cite{Ma-book}.

\section{Scaling of the energy density}
\label{enesca}

In this section we analyze the
out-of-equilibrium behavior of the energy density along the critical
relaxational flow, associated with the quench protocol outlined in
Sec.~\ref{relprot}, within an out-of-equilibrium FSS framework. See
Ref.~\cite{PV-24} for more details.

\subsection{Equilibrium finite-size scaling}
\label{equfss}

In the absence of external symmetry-breaking fields, the equilibrium
energy density $E_e$ is expected to behave as~\cite{PV-02}
\begin{eqnarray}
E_e(K,L) \approx E_{\rm reg}(r) + L^{-(d-y_r)} {\cal E}_e(\Upsilon),
  \label{leadINGFSSene}
\end{eqnarray}
where $d$ is the dimension of the system, and
\begin{equation}
  r \equiv K_c - K,\quad \Upsilon = r \,L^{y_r},\quad y_r=1/\nu.
  \label{Upsilondef}
\end{equation}
The first contribution $E_{\rm reg}(r)$ is a regular function of the
deviation $r$ from the critical point.  The second scaling term
represents the nonanalytic contribution from the critical modes, and
the scaling function ${\cal E}_e({\Upsilon})$ is expected to be
universal, apart from a multiplicative constant and a normalization of
$\Upsilon$. Note, however, that the scaling term turns out to be
subleading with respect to the regular one, due to the fact that
$d-y_r>0$ at continuous transitions (if $y_r\ge d$ the singularity
becomes inconsistent with a continuous transition). Therefore, the
behavior of the energy density is dominated by the analytical
background, which is associated with a mixing with the identity
operator in the field-theoretical setting.

For later applications, it is useful to focus on the subtracted energy
density
\begin{eqnarray}
&&  E_{se}(r,L) \equiv E_e(r,L) - E_c,\label{diffe}\\
&&E_c \equiv E_e(r=0,L\to\infty) = E_{\rm reg}(0).
\nonumber
\end{eqnarray}
Therefore $E_{se}(r=0,L\to\infty) = 0$ by definition.  A precise
estimate of $E_c$ has been already reported in Eq.~(\ref{ecrit}).  The
scaling behavior of the subtracted energy density $E_{se}$ is
trivially obtained from Eq.~(\ref{leadINGFSSene}),
\begin{eqnarray}
  &&E_{se}(r,L) \approx \Delta E_{\rm reg}(r) + L^{-(d-y_r)}
  {\cal E}_e(\Upsilon),
  \label{leadINGFSSenesub}\\
&&\Delta E_{\rm reg}(r)\equiv E_{\rm reg}(r) - E_{\rm reg}(0) = b_1 \,r +
  O(r^2),\qquad
\nonumber
\end{eqnarray}
for generic systems ($b_1$ is a nonuniversal constant).  Note that in
the FSS limit, where $r\sim L^{-y_r}$, the regular term of the
subtracted energy density $E_{se}(r,L)$ is subleading if
\begin{eqnarray}
  2 y_r - d = \frac{2-d\nu}{\nu} = \frac{\alpha}{\nu} > 0,
\label{relcond}
\end{eqnarray}
where $\alpha$ is the specific-heat exponent. This is the case of the
3D Ising universality class, where $\alpha\approx 0.110$. Therefore,
for the 3D Ising models the contribution of the regular term is
subleading. However, it gives rise to very slowly decaying
$O(L^{-\alpha/\nu})$ scaling corrections, with $\alpha/\nu\approx
0.1747$, which make the observation of the asymptotic equilibrium FSS
of $E_{se}(r,L)$ a hard task in numerical analyses.  As we shall see,
this problem gets overcome along the critical relaxational flow, where
the contributions from the regular term are suppressed, and the
scaling corrections turn out to be $O(L^{-\omega})$ with
$\omega\approx 0.83$.

\subsection{Out-of-equilibrium finite-size scaling}
\label{outfss}

To monitor the out-of-equilibrium behavior along the critical
relaxational flow outlined in Sec.~\ref{relprot}, we consider the
subtracted post-quench energy density
\begin{eqnarray}
  E_s(t,r,L) \equiv E(t,r,L)-E_c,
  \label{diffet}
\end{eqnarray}
where $E_c$ is the equilibrium critical value of the energy density in the
thermodynamic limit, cf. Eq.~(\ref{ecrit}).  Asymptotically, due to
the large-time thermalization at the critical point (guaranteed by the
relaxational flow for sufficiently large times at finite volume),
$E_s$ is suppressed as $E_s(t\to\infty)\sim L^{-(d-y_r)}$.

The out-of-equilibrium FSS of $E_s$ can be obtained by extending the
equilibrium FSS discussed in Sec.~\ref{equfss}, adding a further
dependence on the time scaling variable~\cite{PRV-18,RV-21,TV-22}
\begin{equation}
 \Theta = t\,L^{-z}.\label{Thetadef}
\end{equation}
Therefore, the out-of-equilibrium FSS of the subtracted energy density
$E_s$ is expected to behave as~\cite{PV-24}
\begin{eqnarray}
  \Omega(t,r,L) \equiv  L^{d-y_r} E_s(t,r,L) \approx {\cal
    A}(\Theta,\Upsilon),
  \label{erscal}
\end{eqnarray}
where ${\cal A}$ is an out-of-equilibrium scaling function,
which is expected to be universal apart from a multiplicative constant
and trivial normalizations of the arguments.

Note that for Ising-like systems, the above asymptotic scaling
behavior is expected to hold for any $\Theta\ge 0$.  However, the
scaling corrections are not uniform for $\Theta\to 0$.  Indeed, as
conjectured and numerically verified in Ref.~\cite{PV-24} (see also
Ref.~\cite{RV-24} for similar results within quantum transitions and
the unitary quantum dynamics), the contribution of the regular term to
the subtracted energy density gets rapidly suppressed along the
critical relaxational flow. This is related to the fact that the
effective thermalization of the short-ranged modes that give rise to
the regular background of the energy density,
cf. Eq.~(\ref{leadINGFSSene}), requires a much shorter time scale
$\tau_B$ than the critical modes, i.e., $\tau_B\ll \tau\sim L^z$, so
that their asymptotic equilibrium contribution $E_c$ is effectively
approached instantaneously in terms of the rescaled time variable
$\Theta=t/L^z$~\cite{PV-24}. Thus the residual $O(L^{-n\alpha/\nu})$
equilibrium contributions ($n\ge 1$ is an integer number), arising
from the expansion of the regular term in powers of $r$,
cf. Eq.~(\ref{leadINGFSSenesub}), are expected to disappear from the
asymptotic out-of-equilibrium FSS of the subtracted energy density at
any $\Theta>0$.  Therefore the scaling corrections are expected to be
$O(L^{-\omega})$ for any fixed $\Theta>0$, with $\omega\approx 0.83$
significantly larger than $\alpha/\nu\approx 0.17$.  This scenario
will be further confirmed by the numerical analyses presented in the
next section.  This is a crucial point for the effectiveness of our
approach based on the analysis of the subtracted energy density along
the relaxational flow, and, in particular, to obtain an accurate
estimate of the dynamic exponent $z$.

However, assuming that the limit $\Theta\to 0$ matches with the time
dependence of the subtracted energy density (which must be smooth in
finite-size systems), the disappearance of the $O(L^{-\alpha/\nu})$
scaling corrections from the asymptotic behaviors at fixed $\Theta>0$
must reemerge as a nonanalytical behavior of the scaling function
${\cal A}(\Theta,\Upsilon)$ in the $\Theta\to 0$ limit,
i.e.,~\cite{PV-24}
\begin{equation}
  {\cal A}(\Theta,\Upsilon) \approx f_0(\Upsilon) +
  f_1(\Upsilon) \, \Theta^{\alpha/\nu}.
  \label{thetasmall}
\end{equation}

In conclusion, for $\Theta>0$ we generally expect
\begin{eqnarray}
  &&\Omega(t,r,L) = {\cal A}(\Theta,\Upsilon) +
  \label{generalansatz}\\
&&\quad +  L^{-\omega} {\cal
    A}_{\omega}(\Theta,\Upsilon) + ... +
  L^{-\omega_2} {\cal
    A}_{\omega_2}(\Theta,\Upsilon) + ...
\nonumber
\end{eqnarray}
where the dots indicate more suppressed terms, such as
$O(L^{-n\omega})$ with $n>1$, and corrections from other irrelevant RG
perturbations.  The subleading scaling corrections associated with
$\omega_2\approx 2.02$ arise from the breaking of the rotational
invariance in cubic-lattice
systems~\cite{CPRV-98,SD-17,Hasenbusch-21}.

We finally note that the subtraction of the critical energy density
$E_c$ can be avoided by evaluating the derivative of the energy
density along the relaxational flow. Indeed, by taking the time
derivative of $\Omega(t,r,L)$ defined in Eq.~(\ref{erscal}), we obtain
the relation
\begin{eqnarray}
L^z {d\Omega(t,r,L)\over dt} &=&
L^{d-y_r+z} {d E(t,r,L)\over dt}  \label{dererscal}\\
&\approx& {\partial {\cal
      A}(\Theta,\Upsilon)\over \partial\Theta}
= {\cal A}_t(\Theta,\Upsilon),
\nonumber
\end{eqnarray}
where ${\cal A}_t$ is another scaling function whose computation does
not require the knowledge of $E_c$.
  
\section{Numerical results}
\label{numres}

\begin{figure}[tbp]
\includegraphics[width=0.9\columnwidth, clip]{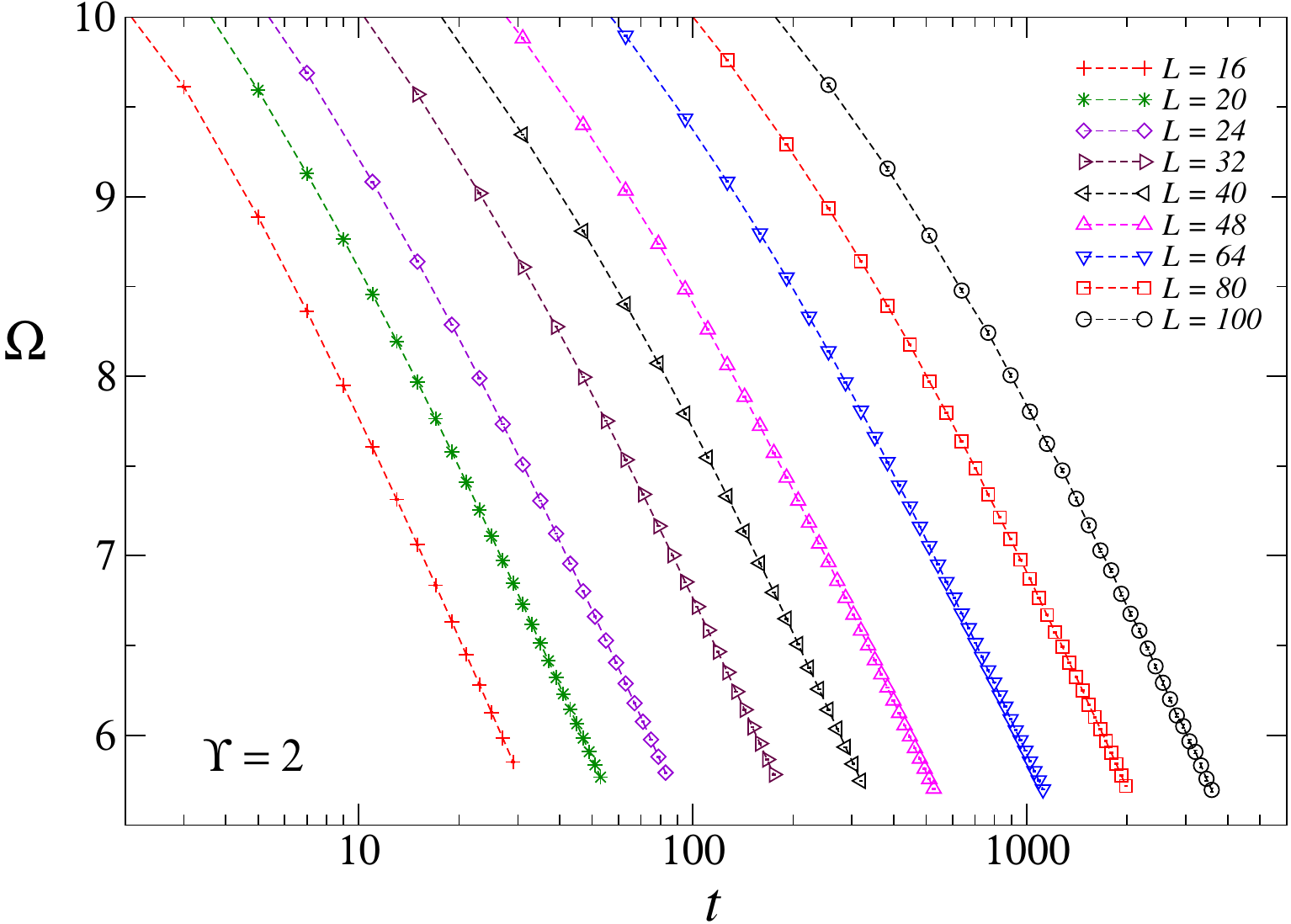}
\caption{Data of $\Omega\equiv E_s L^{d-y_r}$ versus the MC time $t$
  for various lattice sizes up to $L=100$, keeping $\Upsilon=2$ fixed
  (note that we use a logarithmic scale for the time axis).  The
  statistical errors of the data are very small and hardly visible.}
\label{rawdataY2}
\end{figure}

In this section we present our numerical analyses of the
out-of-equilibrium behavior of the subtracted energy density
$E_s(t,r,L)$, cf. Eq.~(\ref{diffet}), along the critical relaxational
flow driven by standard single-spin-flip Metropolis upgradings: the
transition probability for a flip of the link variable $\sigma_{{\bm
    x},\mu}$ is given by
\begin{equation}
  P(\sigma_{{\bm x},\mu}\to -\sigma_{{\bm x},\mu}) =
  {\rm Min}[1,e^{-\Delta H_G}],
    \label{metroupg}
    \end{equation}
      where $\Delta H_G$ is the variation of the lattice Hamiltonian
      (\ref{Hgz2}) when replacing $\sigma_{{\bm x},\mu}\to
      -\sigma_{{\bm x},\mu}$. One MC time unit corresponds to a global
      sweep of a single Metropolis upgrading proposal for each bond
      variable (we use a checkerboard upgrading scheme, where we
      separately upgrade link variables along a given direction at odd
      and even sites). Its acceptance ratio turns out to be relatively
      small at the critical point $K_c$, about 2\% [note that this
        acceptance ratio is much smaller than that of the analogous
        Metropolis upgrading at the critical point $J_c$ of the Ising
        model (\ref{His}), which is about 50\%]. We consider
      cubic-like systems of size $L$ and volume $V=L^3$, with periodic
      boundary conditions.

We present results for the critical relaxational flow of the 3D
${\mathbb Z}_2$ gauge model at some fixed values of $\Upsilon$, i.e.,
$\Upsilon=2,4,8$, up to lattice sizes $L=100$ (more precisely for
$L=16,\,20,\,24,\,32,\,40,\,48,\,64,\,80,\,100$), obtained by
averaging the time-dependent energy density defined in
Eq.~(\ref{defened}), over a large number of trajectories implementing
the critical relaxational flow.  In practice, along the equilibrium
run at $K<K_c$, corresponding to a given value of $\Upsilon$, we start
a trajectory every $n \approx 0.2 \,L^z$ sweeps at equilibrium, where
we used the preliminary value $z=2.6$ (this distance between
trajectories provides a reasonable compromise to get almost
decorrelated starting configurations, as checked by preliminary
tests).  We collect about $O(10^6)$ relaxational trajectories for
$L\le 80$, and about $2.5 \times 10^5$ trajectories for the largest
lattice size $L=100$.  Any trajectory typically runs up to $t\approx
0.02 \, L^z$, thus increasing its duration when increasing $L$, to
obtain averages at fixed $\Theta=t/L^z$ up to $\Theta\approx 0.02$.
Therefore the numerical effort per trajectory increases as $L^{3+z}$
with increasing $L$.  The error on the average of the energy density
over the trajectories is estimated by a blocking procedure to properly
take into account the residual autocorrelations among the sequential
trajectories along the equilibrium run.  The large number of
trajectories allows us to estimate the average energy density along
the critical relaxational flow with high accuracy, for example with a
relative accuracy of about $2 \times 10^{-6}$ for $L=100$.

In Fig.~\ref{rawdataY2} we show some raw data of the subtracted
energy density $E_s$, in particular $\Omega= E_s L^{d-y_r}$, versus
the time $t$, for various values of $L$ at fixed $\Upsilon=r
L^{y_r}=2$. As we shall see, the analysis of these data within the
out-of-equilibrium FSS framework outlined in Sec.~\ref{enesca} allows
us to estimate the dynamic exponent $z$ associated with the purely
relaxational dynamics, by determining the optimal collapse of the data
when plotting them versus $\Theta=t L^{-z}$.

Since the data along the critical relaxational flow are highly
correlated, we must pay some attention to extract a reliable estimate
of the power law controlling the time dependence.  For this purpose,
it is convenient to consider the time $t(\Omega)$ as a function of the
rescaled subtracted energy density $\Omega$ (this is well defined
because $\Omega$ is a monotonic function of $t$ along the critical
relaxational flow).  Numerically, this can be easily estimated by
linearly interpolating the data of $\Omega(t,r,L)$ as a function of
$t$ (one may also use higher-order interpolations), using a
straightforward propagation of the statistical error.  One can easily
derive the corresponding out-of-equilibrium FSS from that of the
subtracted energy density, cf. Eq.~(\ref{generalansatz}), obtaining
\begin{eqnarray}
&&t(\Omega,r,L) \approx L^{z} 
    F(\Omega,\Upsilon) +   \label{generalansatzt}\\
&&\quad +     L^{z-\omega} F_{\omega}(\Omega,\Upsilon) + ... + 
  L^{z-\omega_2} F_{\omega_2}(\Omega,\Upsilon) + ...
\nonumber
\end{eqnarray}
In our analyses we also consider the difference
\begin{eqnarray}
  \Delta t(\Omega,r_2,r_1,L) &\equiv&
  t(\Omega,r_2,L) - t(\Omega,r_1,L)   \label{difft}\\
  &\approx& L^z
  F_{d}(\Omega,\Upsilon_2,\Upsilon_1),
\nonumber
\end{eqnarray}
where $\Upsilon_i = r_i L^{y_r}$, because we noted that these differences are
less affected by scaling corrections.

\begin{figure}[tbp]
\includegraphics[width=0.9\columnwidth, clip]{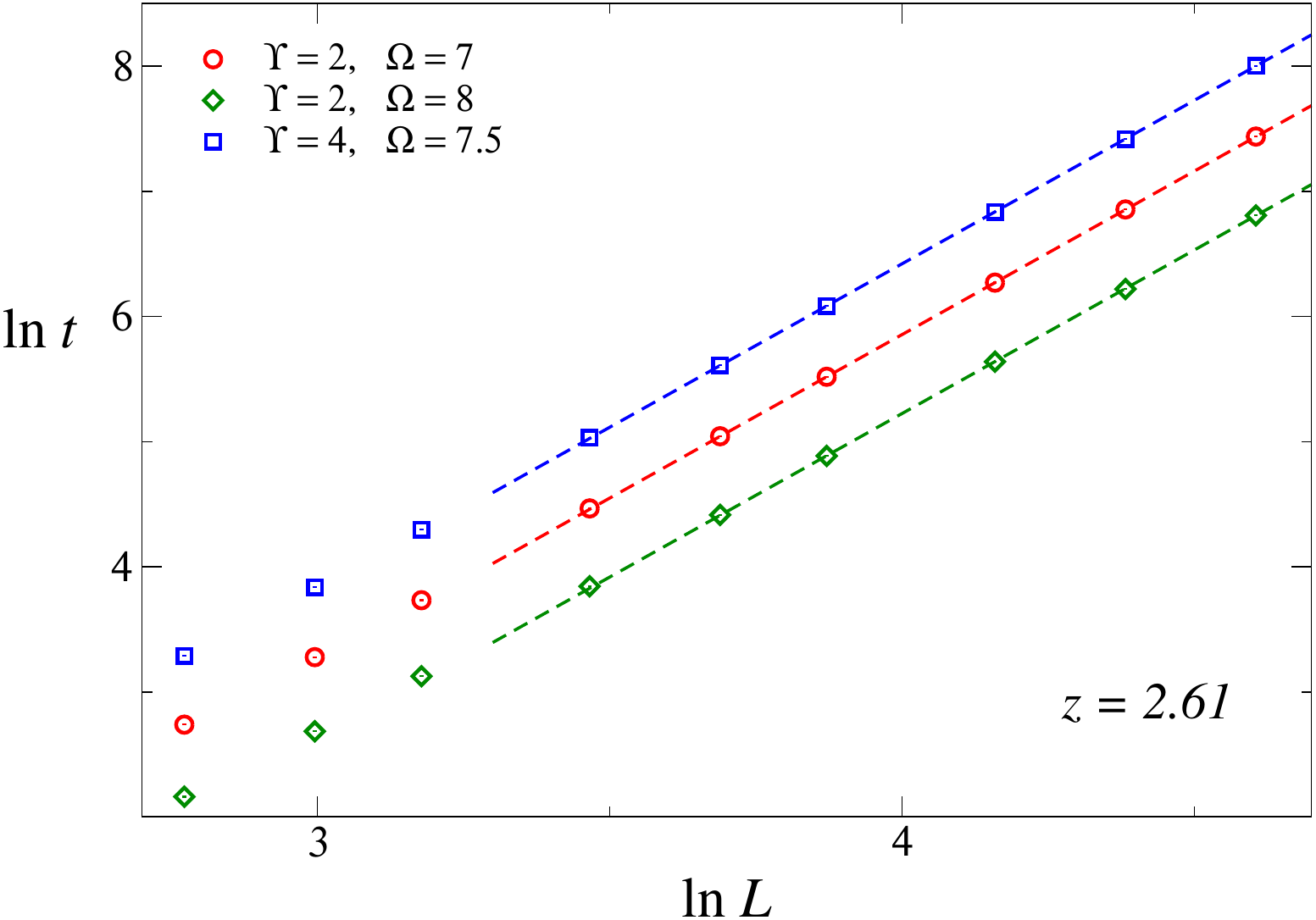}
\caption{Some data of the time versus the lattice size, computed at
  fixed $\Upsilon$ and $\Omega$, in logarithmic scale. The lines show
  linear fits of the data for $L\ge 40$ with $z=2.61$ (with
  $\chi^2/{\rm d.o.f}\lesssim 1$). The statistical errors of the data
  are hardly visible in these plots.  }
\label{tomega}
\end{figure}

To avoid the problems of the statistical correlations among the data
for different values of $\Omega$, we perform fits of the data of
$t(\Omega,r,L)$ at fixed values of $\Upsilon$ and $\Omega$, so that
the data considered are statistically independent.  We consider, and
compare the results, of fits to various ansatz functions, such as
\begin{eqnarray}
  &(a): \;  &t(\Omega,\Upsilon,L)=a_0 \, L^z,\label{fit1}\\
  &(b): \;   &t(\Omega,\Upsilon,L)=a_0 \,L^z \left(1
  + a_1 \,L^{-\omega}\right), \label{fit2}\\
  &(c): \;   &t(\Omega,\Upsilon,L)=a_0 \, L^{z}\left(1 + a_1 \,L^{-\omega}
  + a_2\, L^{-2}\right).\qquad \label{fit3}
  \end{eqnarray}
Of course, the coefficients $a_i$ of the above asymptotic behaviors
depend on $\Upsilon$ and $\Omega$. The fit (c) is also useful because
for the smallest lattice sizes the scaling corrections are apparently
dominated by contributions approximately decaying as $O(L^{-2})$,
which may, for example, be associated with the next-to-leading
irrelevant RG perturbations associated with the lattice anisotropy,
cf.  Eq.~(\ref{generalansatz}).  In Table \ref{fits}, we report the
results of some of the above fits, at some values of $\Omega$ that
turn out to be optimal. For example, the region $\Omega\approx 7$
provides the most stable fits for $\Upsilon=2$, corresponding to
$\Theta=t/L^z\approx 0.01$. Fig.~\ref{tomega} shows some data of the
time $t(\Omega,\Upsilon,L)$ for some values of $\Omega$ and
$\Upsilon$, with their best fits, to highlight the accuracy of the
estimate of $z$. We also note that the results are subject to larger
scaling corrections at larger values of $\Omega$, corresponding to
smaller values of $\Theta$, and provide substantially consistent, but
less accurate, results. This is easily explained by the expected
crossover behavior of the scaling corrections in the limit $\Theta\to
0$, where they are expected to pass from $O(L^{-\omega})$ at
$\Theta>0$ to $O(L^{-\alpha/\nu})$ for $\Theta=0$.

\begin{table}
\begin{tabular}{cccclc}
  \hline\hline
  & $\quad \Omega \quad$ & Ansatz &
  $\quad L_{\rm min}\quad $ & $\quad z$ & $\chi^2/{\rm
    d.o.f.}$ \\ \hline
$\Upsilon=2$  &  6  &  (a)  &  40 & 2.613(1)  & 0.9  \\
  &    &    &  48 & 2.613(2)  & 1.3  \\

    &  7  &  (a)  &  40 & 2.610(2)  & 0.5  \\
  &    &    &  48 & 2.611(2)  & 0.3  \\

  &    &  (b)  &  40 & 2.617(15) & 0.6 \\
  
  &    &  (c)  &  16 & 2.589(10) & 0.6 \\
  &    &       &  20 & 2.591(15) & 0.4  \\
  &    &       &  24 & 2.594(24) & 0.4    \\

  &  8  &  (a)  &  40 & 2.607(2)  & 2.5  \\
  &     &       &  48 & 2.612(3)  & 0.8  \\\hline  

$\Upsilon=4$  &  7.5  &  (a)  &  40 & 2.608(1)  & 0.4 \\
  &    &    &  48 & 2.608(1)  & 0.4   \\
  &    &  (c)  &  16 & 2.611(5)  & 1.2   \\
    &    &    &  20 & 2.613(8)  & 1.3   \\\hline  

$D_{24}$ & 7.5 &  (a)   & 32  &  2.603(1)  & 1.0  \\
        &  &           & 40  &  2.606(2)  & 0.4  \\
        &  &           & 48  &  2.605(3)  & 0.3  \\ 
  &  &     (c)   & 16  &  2.598(13) & 0.4  \\
    &  &        & 20  &  2.600(20) & 0.4  \\ \hline

$D_{48}$ & 8 &  (a)     & 48  &  2.592(6)  & 1.5  \\
        &  &           & 64  &  2.600(11)  & 1.6 \\\hline

 $C_{24}$  & 7.5 &  (a)   & 40  &  2.608(1)  & 0.8  \\
        &  &           & 48  &  2.609(1)  & 0.7  \\
  &  &           & 64  &  2.610(2)  & 0.3  \\
  & & (b)  & 40 & 2.618(8) &  0.5 \\
  & &   & 48 & 2.616(13) &  0.4 \\\hline

 $C_{248}$  & 8 &  (a)   & 48  &  2.610(1)  & 4.9  \\
        &  &           & 64  &  2.613(1)  & 2.8  \\
  \hline\hline
\end{tabular}
\caption{ Some results of fits of the data of $t(\Omega,r,L)$ at fixed
  values of $\Omega$, cf. Eq.~(\ref{generalansatzt}) for $\Upsilon=2$,
  $\Upsilon=4$, the difference between the data for different
  $\Upsilon$ at fixed $\Omega$ (denoted by $D_{24}$ and $D_{48}$), and
  the combined fit of the data for $\Upsilon=2$, $\Upsilon=4$ (denoted
  by $C_{24}$) and those for $\Upsilon=2,\,4,\,8$ (denoted by
  $C_{248}$). We report results for fits of type (a), (b), and (c)
  defined in Eqs.~(\ref{fit1})-(\ref{fit3}).  The values of $\Omega$
  chosen for the fits correspond to the optimal region
  $\Theta=t/L^z\approx 0.01$, where scaling corrections turn out to be
  smaller. On the basis of these results, we arrive at $z=2.610(15)$
  as our final estimate, where the error takes into account the small
  differences of the results, including most results reported in the
  table (the error is essentially dominated by the differences of the
  various fits, while statistical errors are generally much smaller).
  Other fits for different values of $\Omega$ give consistent, but
  less precise, results.}
  \label{fits}
\end{table}

\begin{figure}[tbp]
  \includegraphics[width=0.9\columnwidth, clip]{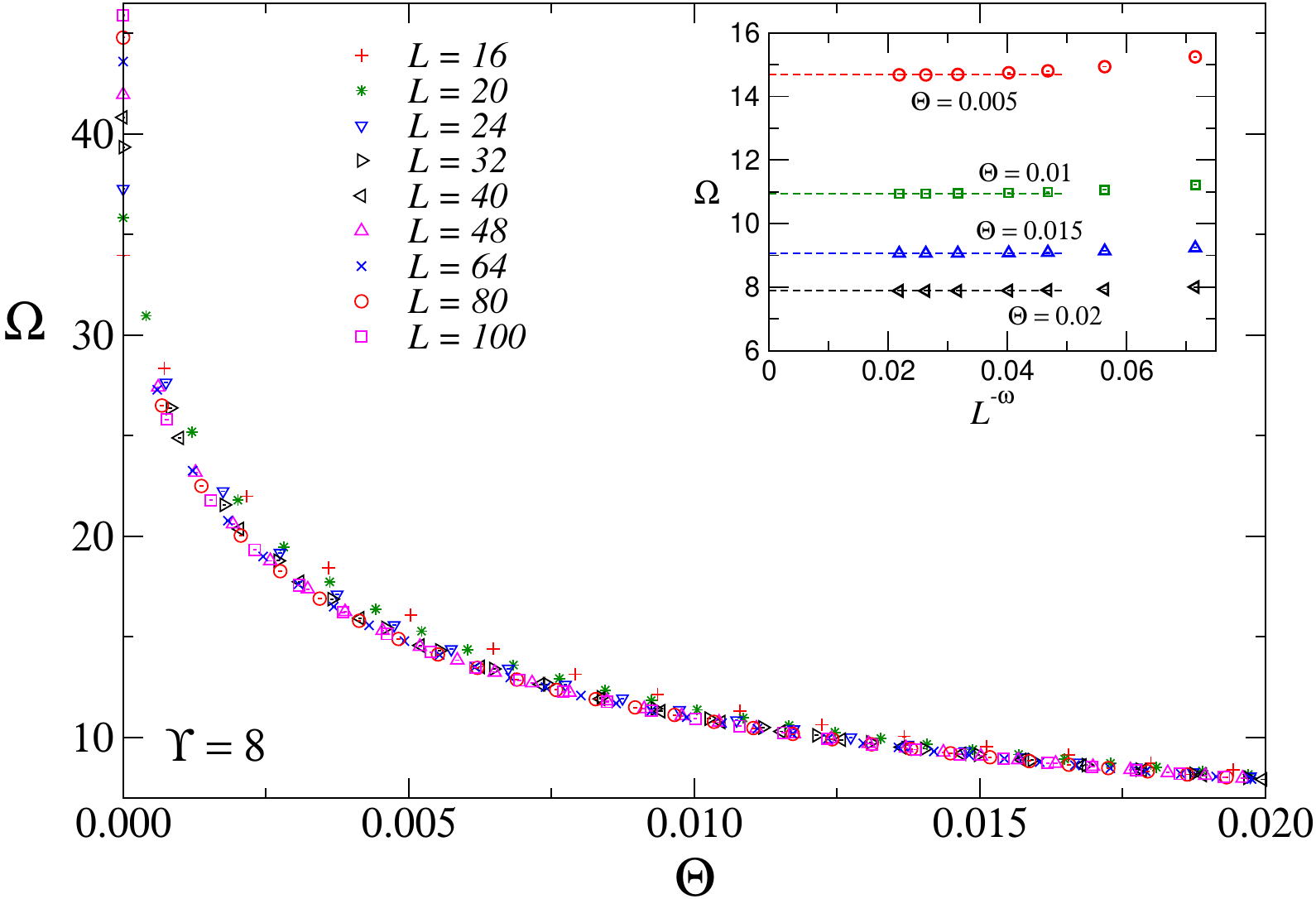}
  \includegraphics[width=0.9\columnwidth, clip]{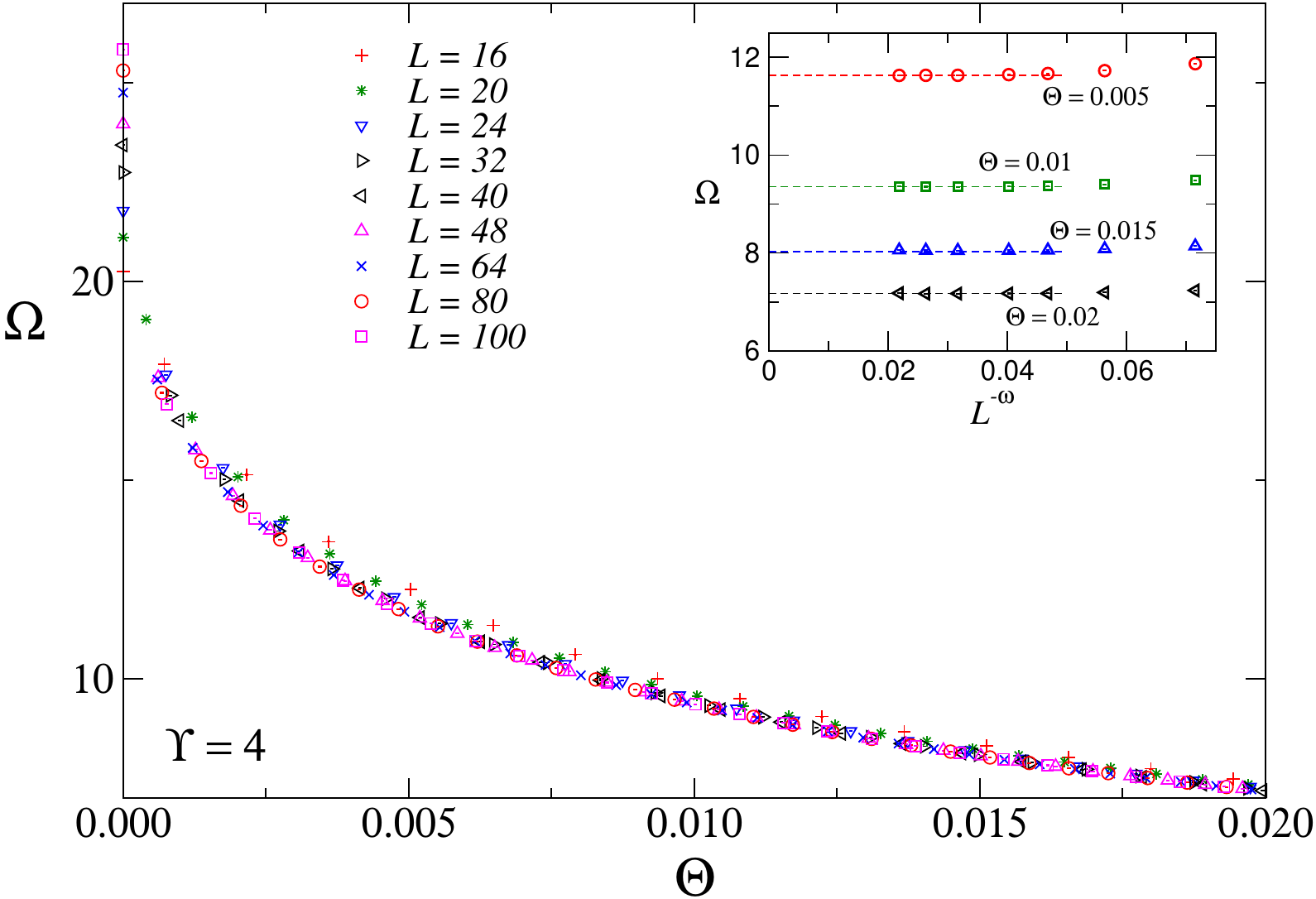}
  \includegraphics[width=0.9\columnwidth, clip]{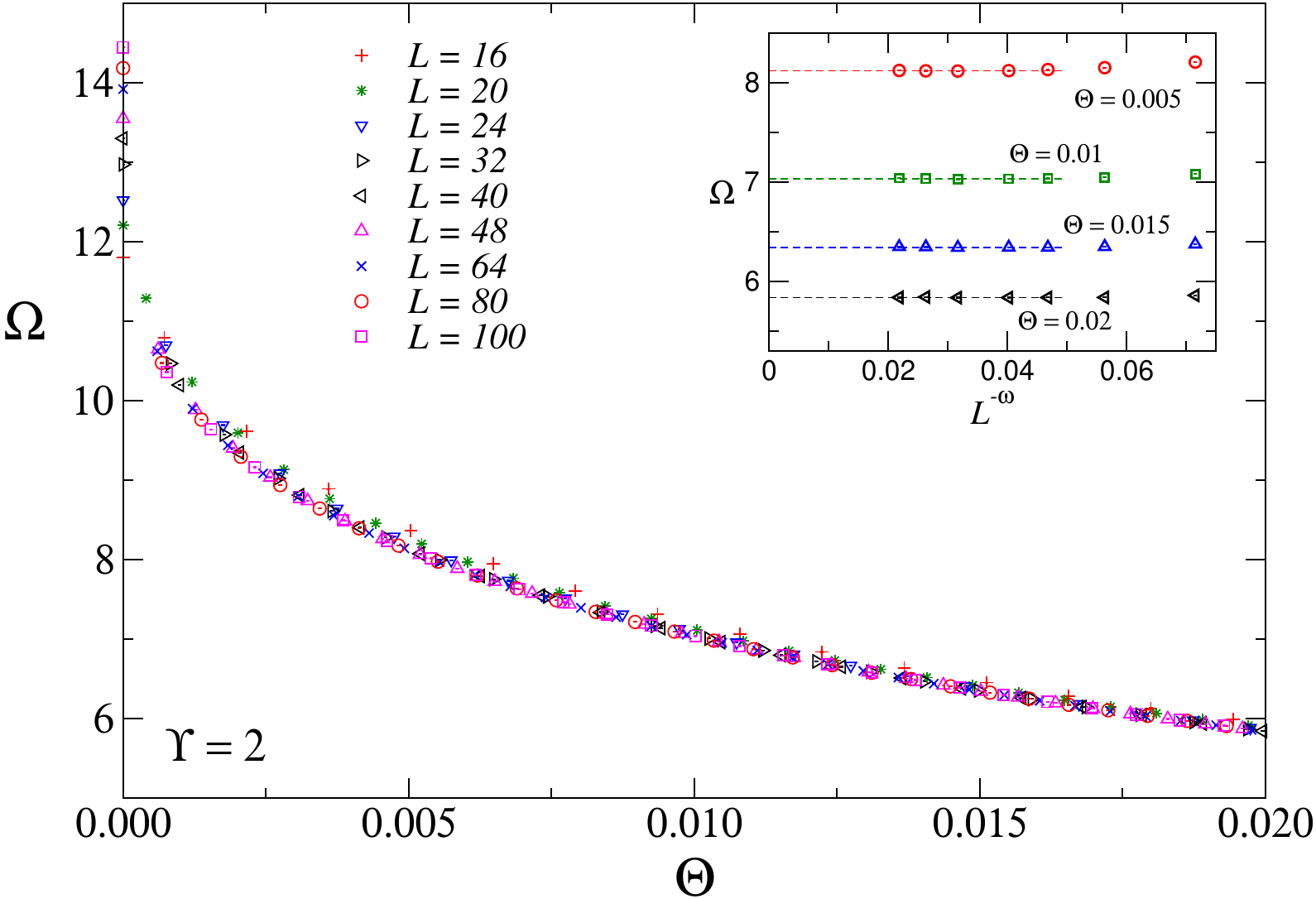}
  \caption{We plot $\Omega\equiv E_s L^{d-y_r}$ versus $\Theta=t/L^z$
    using our optimal estimate $z=2.61$, for $\Upsilon=2$ (bottom),
    $\Upsilon=4$ (middle), and $\Upsilon=8$ (top).  The insets show
    the convergence to the asymptotic scaling behavior for some values
    of $\Theta$, plotting the data of $\Omega$ versus $L^{-\omega}$
    which is the expected rate of convergence, while the
    dashed lines represent the asymptotic value ${\cal
      A}(\Theta,\Upsilon)$. }
    \label{OmegavsTheta}
\end{figure}

On the basis of the results obtained by the various fits, see in
particular Table~\ref{fits}, we report
\begin{equation}
  z = 2.610(15)
  \label{zestimate}
\end{equation}
as our final estimate, where the error takes into account the range of
results obtained by the different fits considered, and also the
variation due to uncertainty on the available estimates of $K_c$ and
$E_c$, cf. Eqs.~(\ref{z2gaugecr}) and (\ref{ecrit}) respectively,
which are much smaller.  The quality of the resulting scaling behavior
within the whole critical relaxational flow can be appreciated looking
at the plots shown in Fig.~\ref{OmegavsTheta}, where we report
$\Omega(t,r,L)$ versus $\Theta$ using the estimate (\ref{zestimate}):
the asymptotic collapse of the curves for any $\Theta>0$ is clearly
observed. It is also worth mentioning that consistent, although less
precise, results are obtained by looking for the optimal collapse of
the data of the time derivative of $\Omega$,
cf. Eq.~(\ref{dererscal}), which does not require the knowledge of
$E_c$.

\begin{figure}[tbp]
  \includegraphics[width=0.9\columnwidth, clip]{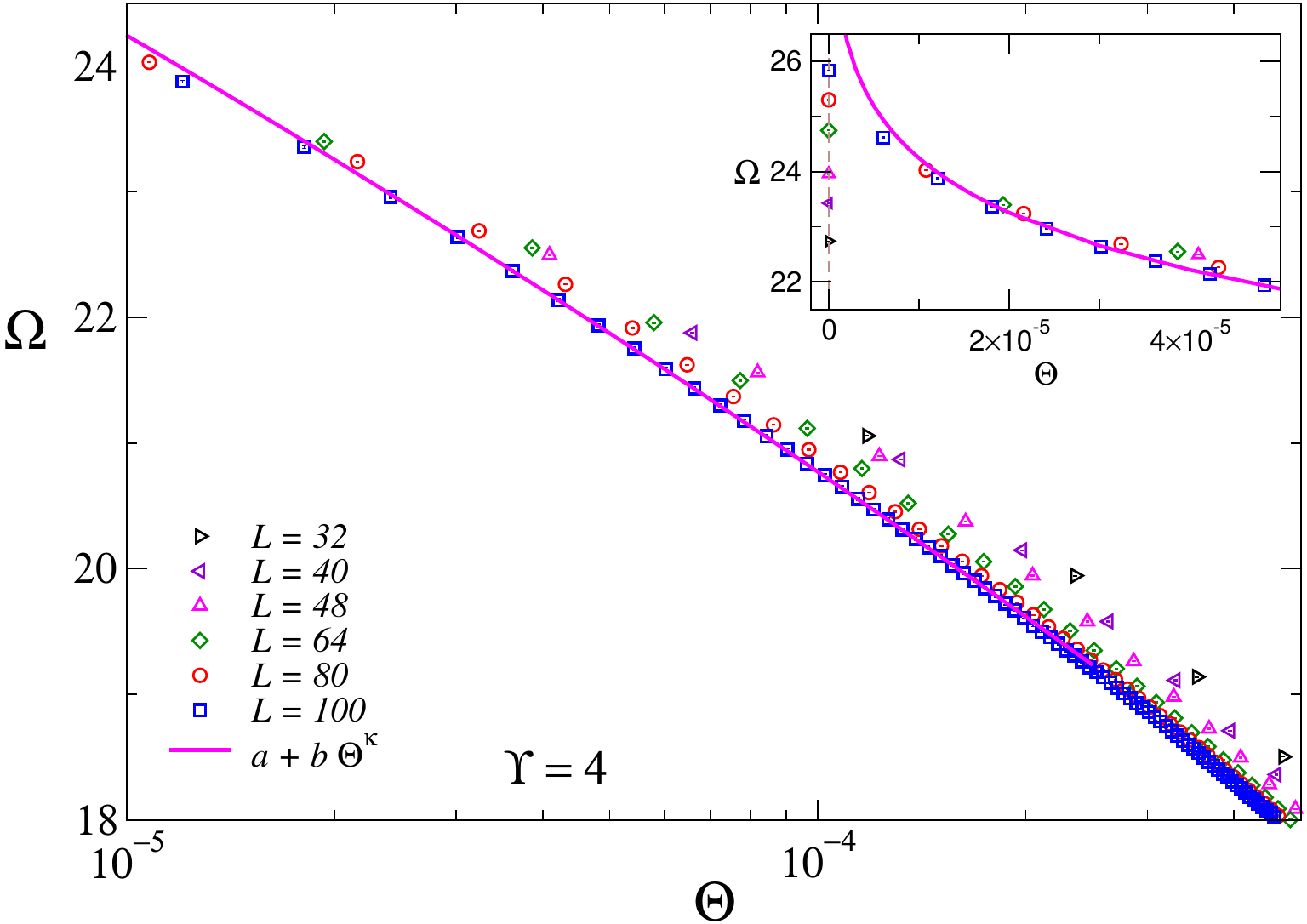}
    \includegraphics[width=0.9\columnwidth, clip]{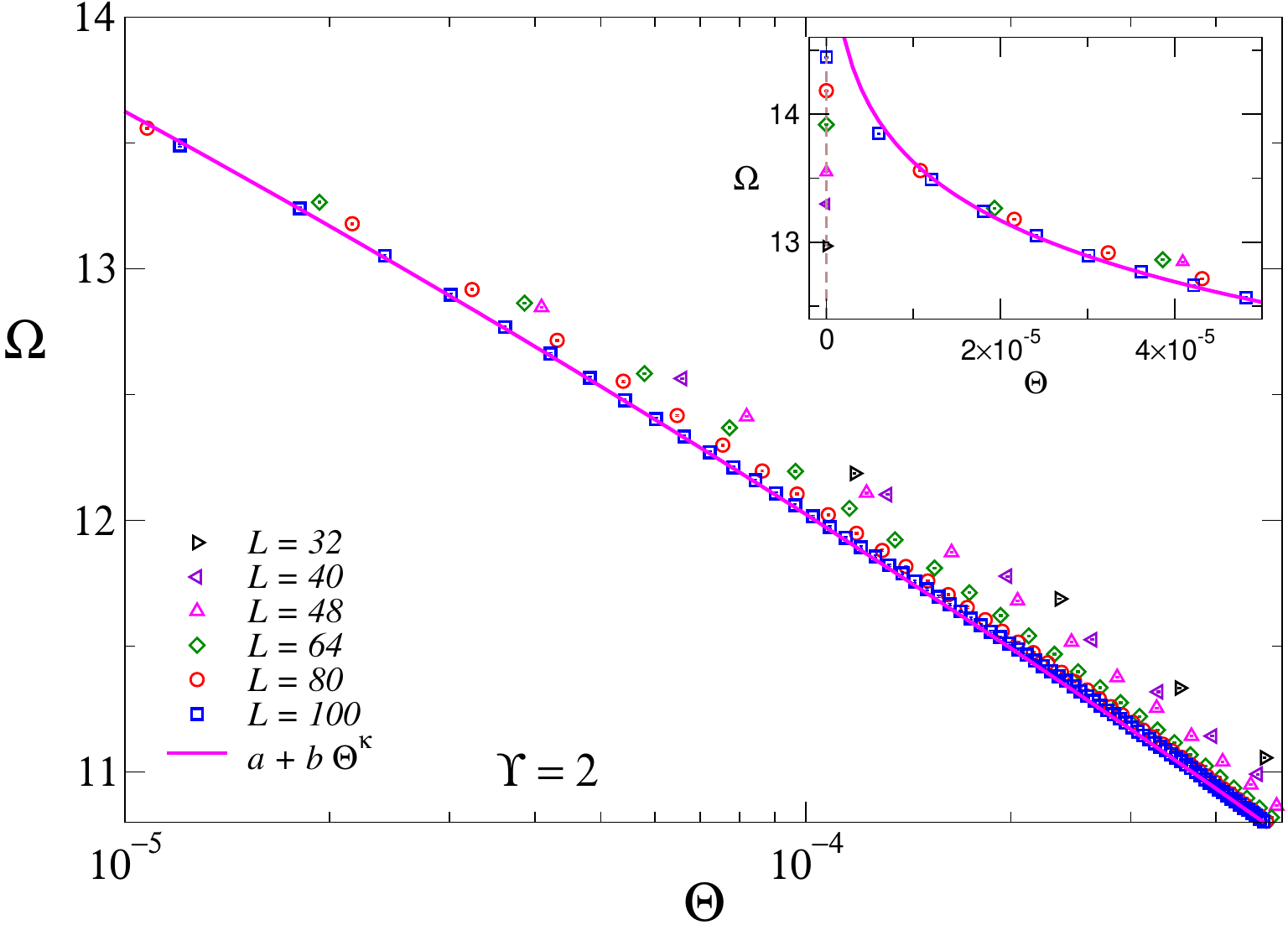}
  \caption{We show data of $\Omega(t,r,L)$ for small values of
    $\Theta=t/L^z$ using our optimal estimate $z=2.61$, at fixed
    $\Upsilon=2$ (bottom) and $\Upsilon=4$ (top).  The data support
    the expected behavior (\ref{thetasmall}). This is shown by the
    full line, corresponding to a fit of the $L=100$ data in the range
    $3\times 10^{-5}\lesssim \Theta\lesssim 3 \times 10^{-4}$ to
    $a+b\Theta^\kappa$ with $\kappa = \alpha/(\nu z)\approx 0.0669$
    (we prefer to show this fit assuming $L=100$ large enough, instead
    of extrapolating the data to $L\to\infty$, which requires a quite
    cumbersome procedure in the small-$\Theta$ crossover region
    $\Theta\lesssim 10^{-4}$). The insets show the same data close to
    $\Theta=0$ without logarithmic scale (the vertical dashed lines indicate
    the $\Theta=0$ starting point of the relaxational flow), to
    highlight the crossover from the equilibrium behavior at $t=0$ to
    the out-of-equilibrium scaling behavior at small values
    of~$\Theta$.}
\label{smalltheta}
\end{figure}

\begin{figure}[tbp]
  \includegraphics[width=0.9\columnwidth, clip]{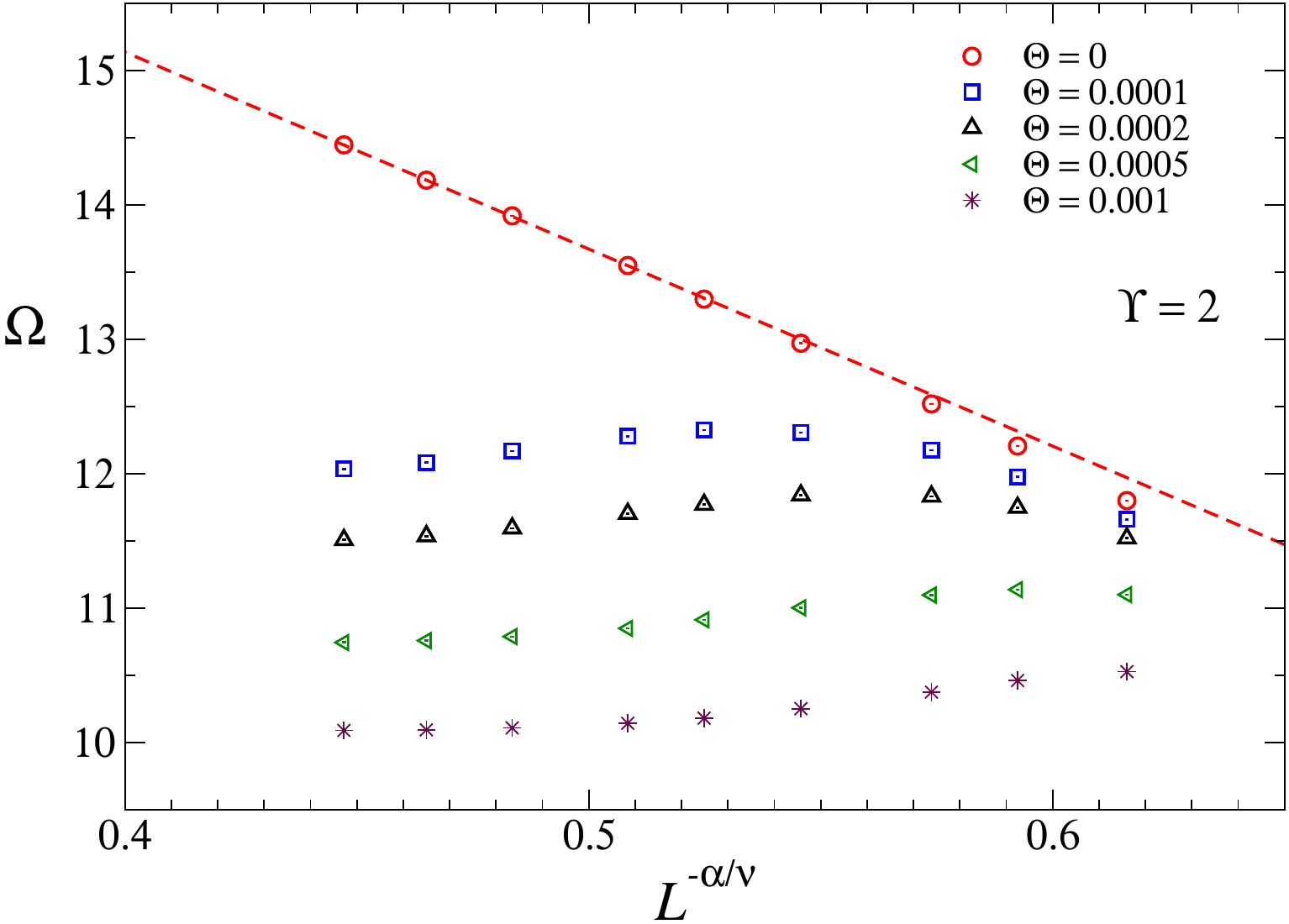}
  \caption{We show data of $\Omega(t,r,L)$ for some values of
    $\Theta=t/L^z$, from $\Theta=0$ to $\Theta = 0.01$, using our
    optimal estimate $z=2.61$, at fixed $\Upsilon=2$. They are plotted
    versus $L^{-\alpha/\nu}$ which is the expected rate of suppression
    of the leading scaling corrections at $t=\Theta=0$, as confirmed
    by the almost linear behavior of the equilibrium data for
    $\Theta=0$ (the dashed line shows a linear fit of the equilibrium
    $t=0$ data for the largest lattice sizes with $L\ge 48$ to $a + b
    L^{-\alpha/\nu}$). Note also the substantially different behavior
    of the scaling corrections for $\Theta>0$ along the critical
    relaxational flow, where the data confirm that the
    $O(L^{-\alpha/\nu})$ corrections vanish at fixed $\Theta>0$,
    leaving only more suppressed $O(L^{-\omega})$ corrections.  }
\label{scalingcorr}
\end{figure}

We also note that the behavior for small $\Theta$ confirms the
$\Theta\to 0$ nonanalyticity (\ref{thetasmall}) of the scaling
function ${\cal A}(\Theta,\Upsilon)$, see Fig.~\ref{smalltheta} where
data for $\Upsilon=2$ and $\Upsilon =4$ are shown. Note also that the
data at $t=0$, or $\Theta=0$, show clearly a different approach to the
asymptotic FSS limit with respect to the data at $\Theta>0$.  Indeed,
while at $t=0$ the equilibrium FSS is approached from below, at
fixed $\Theta>0$ the out-of-equilibrium FSS is approached from above,
see, e.g., the plots reported in Fig.~\ref{scalingcorr}. This is in
agreement with the fact that the leading $O(L^{-\alpha/\nu})$ scaling
corrections arising from the regular term at equilibrium (also
corresponding to the so-called mixing with the identity operator in
the field-theoretical language), cf. Eq.~(\ref{leadINGFSSenesub}),
get suppressed along the critical relaxational flow, whose asymptotic
approach is expected to be characterized by more standard
$O(L^{-\omega})$ decaying corrections~\cite{PV-24,RV-24}.

We finally mention that the out-of-equilibrium estimate
(\ref{zestimate}) improves earlier results obtained by other
approaches, reported in Refs.~\cite{BKKLS-90,XCMCS-18,BPV-25}.  In
particular, it is in agreement with the result $z=2.55(6)$ obtained
analyzing the equilibrium critical dynamics~\cite{BPV-25}.  We also
note that our new result maintains the apparent small discrepancy with
the estimate $z=2.70(3)$ obtained in Ref.~\cite{XCMCS-18} by
analyzing the out-of-equilibrium behavior of the Polyakov lines when
slowly changing the temperature to cross the transition point.

\section{Conclusions}
\label{conclu}

We have studied the out-of-equilibrium critical dynamics of 3D lattice
${\mathbb Z}_2$ gauge models, along critical relaxational flows
arising from instantaneous quenches to the critical point, driven by
purely relaxational local Metropolis upgradings of the link ${\mathbb
  Z}_2$ gauge variables. We monitor the critical dynamics by computing
the local energy density, which is the simplest gauge-invariant
quantity that can be computed in a lattice gauge theory.  We analyze
the behavior along the critical relaxational flow within an
out-of-equilibrium FSS framework, exploiting the suppression of the
regular contributions that generally affect the equilibrium
free-energy density and the energy density at the critical
point~\cite{PV-24}.

By matching the numerical data up to relatively large lattice sizes
with the scaling behavior predicted by the out-of-equilibrium FSS
theory outlined in Sec.~\ref{outfss}, we obtain the estimate
$z=2.610(15)$ for the dynamic exponent associated with the local
purely relaxational dynamics of the 3D ${\mathbb Z}_2$ gauge
universality class. This result significantly improves earlier
computations of the dynamic exponent $z$, see
Refs.~\cite{BKKLS-90,XCMCS-18,BPV-25}, obtained by both equilibrium
and out-equilibrium studies.  In particular, our estimate of $z$ is
agreement with the result $z=2.55(6)$ obtained by analyzing the
equilibrium critical dynamics~\cite{BPV-25}.

Our results confirm that the critical slowing down experienced by the
lattice ${\mathbb Z}_2$ gauge models is stronger than that associated
with the relaxational dynamics of the standard Ising universality
class, whose dynamic exponent $z=2.0245(15)$~\cite{Hasenbusch-20} is
significantly smaller.  This difference can be explained by noting
that the duality relation between the ${\mathbb Z}_2$ gauge and Ising
universality classes is nonlocal. Therefore local relaxational
upgradings act on substantially different modes, giving rise to
different critical relaxational dynamics, and therefore different
values of the dynamic exponents controlling the power law of the
critical slowing down.

We finally remark that the out-of-equilibrium method exploited in this
study, based on the study of the critical relaxational flow of the
local gauge-invariant energy density, is expected to be effective to
address the critical dynamics of any system with gauge symmetries,
whose critical modes at their topological transition can be hardly
monitored due to the absence of a local order parameter.  For example,
one interesting case is the 3D lattice Abelian Higgs model with a
one-component complex scalar field, whose continuous transitions do
not have any local order parameter, analogously to the 3D ${\mathbb
  Z}_2$ gauge model, see, e.g.,
Refs.~\cite{BPV-24-ncAH,BPV-24-decQ2,BPV-24-rev}.

\begin{acknowledgments}
H.P. thanks the University of Pisa for the kind hospitality. H.P. would like to
acknowledge support from the project CONCEPT/0823/0052, implemented under the
program ``THALIA 2021-2027'' and co-funded by the European Union through the
Cyprus Research and Innovation Foundation (RIF). C.B. and E.V. acknowledge
support from project PRIN 2022 ``Emerging gauge theories: critical properties
and quantum dynamics'' (20227JZKWP). 
\end{acknowledgments}

\end{document}